\documentclass[twocolumn,10pt,final,conference]{IEEEtran}

\usepackage{cite}
\usepackage{amsmath}
\usepackage{amssymb}
\usepackage{graphicx,dblfloatfix}
\usepackage{subfigure}
\usepackage{a4wide}
\usepackage{algorithm}
\usepackage{algorithmic}
\usepackage{amsfonts}
\usepackage{verbatim}

\makeatletter
\def\ps@headings{%
\def\@oddhead{\mbox{}\scriptsize\rightmark \hfil \thepage}%
\def\@evenhead{\scriptsize\thepage \hfil \leftmark\mbox{}}%
\def\@oddfoot{}%
\def\@evenfoot{}}
\makeatother
\begin{document}
\title{Preferential Attachment Model with Degree Bound and its Application to Key Predistribution in WSN}

\author{\IEEEauthorblockN{
Sushmita Ruj$^\dag$ and Arindam Pal$^\ddag$}
\IEEEauthorblockA{
$^\dag$ Indian Statistical Institute, Kolkata, India. Email: sush@isical.ac.in \\
$^\ddag$ TCS Innovation Labs, Kolkata, India. Email: arindamp@gmail.com
}}

\maketitle{}

\begin{abstract}
Preferential attachment models have been widely studied in complex networks, because they can explain the formation of many networks like social networks, citation networks, power grids, and biological networks, to name a few. Motivated by the application of key predistribution in wireless sensor networks (WSN), we initiate the study of preferential attachment with degree bound.

Our paper has two important contributions to two different areas. The first is a contribution in the study of complex networks. 
We propose preferential attachment model with degree bound for the first time. 
In the normal preferential attachment model, the degree distribution follows a power law, with many nodes of low degree and
a few nodes of high degree. 
In our scheme, the nodes can have a maximum degree $d_{\max}$, where $d_{\max}$ is an integer chosen according to the application. 
The second is in the security of wireless sensor networks. We propose a new key predistribution scheme based on the above model. 
The important features of this model are that the network is fully connected, it has fewer keys, has larger size of the giant component 
and lower average path length compared with
traditional key predistribution schemes and comparable resilience to random node attacks.  

We argue that in many networks like key predistribution and Internet of Things, having nodes of very high degree will be a bottle-neck in communication. Thus, studying preferential attachment model with degree bound will open up new directions in the study of complex networks, and will have many applications in real world scenarios.
\end{abstract}

\textbf{Keywords}: Complex Networks, Preferential Attachment, Key Predistribution, Degree Distribution, Giant Component. 

\section{Introduction}
Internet of Things (IoT) consists of a network of devices which communicate with each other exchanging data and control information. IoT appears in many areas like smart homes, smart cities, smart grids, vehicular networks, peer-to-peer (P2P) networks, agriculture, health-care, to name only a few. Devices can be of many types and might be resource constrained. Since the devices exchange
large volumes of data, communication and storage overheads should be minimized. 

IoT can be represented by a graph, with devices being represented as nodes and communication links as edges. P2P networks \cite{LCCLS02}, sensor networks \cite{BDHAF09}, smart grids \cite{HWSN13,RP14}, Bluetooth networks \cite{WHC05} have all been modeled as random graphs and have been well studied.

IoT might consist of Wireless Sensor Networks that are widely used in many civilian, commercial, and military applications. Some of these networks are vulnerable to attacks, particularly those in the military domain \cite{PGVD14}. The resource constrained nature of sensors 
restricts the use of public key methods to secure communication.
Key predistribution \cite{EG02} is a widely used symmetric key management technique in which cryptographic keys are
preloaded in sensor networks prior to deployment. 
Sensor nodes find out the common keys using a \emph{shared key discovery} phase. 
Messages are encrypted by the source node using the shared key and decrypted at the receiving node using the same key. 

The number of keys in each node is to be minimized, in order to reduce storage space. 
Direct communication between any pair of nodes help in quick and easy transmission of message. 
As mentioned in \cite{RS13}, transmission and reception requires more power than computing. 
So, the number of hops between nodes should be minimized. 
Intermediate nodes might leak information and nodes are prone to capture or compromise. 

Many key predistribution techniques have been proposed that trade-off storage, computation, communication and resilience.
Some of these techniques are random (\cite{EG02,CPS03}) and some are deterministic \cite{CY04,LS05,RR07}. 
The deterministic techniques often do not scale well and have poor resilience when a large number of nodes are compromised.
The randomized techniques are simple to implement and are highly resilient.
In \cite{EG02}, a large key pool or a set of keys is constructed.
A fixed number of $k$ keys are drawn at random without replacement from this key pool.
In \cite{CPS03}, a node chooses $k$ other nodes at random and communicates with them using unique pairwise keys.

There has been research on the structure of networks, where random key predistribution
have been applied.
Given such predistribution schemes, a \emph{key graph} is constructed in the following way.
The set of vertices is the set of nodes in the network.
An edge exists between two vertices if they are within communication range and share a common key.
It has been shown that the key graph arising from \cite{EG02} is an Erdos-Renyi random graph with parameters 
$p = \frac{\ln n + c}{n}$ ($c$ is a real constant), whereas the key graph of
\cite{CPS03} is a $k$-out graph.
Erdos-Renyi (ER) graphs are random graphs often denoted by $G(n,p)$, where $n$ is the number of nodes and $p$ is the probability that 
there exists an edge between two vertices. 
A $k$-out graph is a random graph in which every vertex has an outdegree $k$, where the $k$ vertices are chosen uniformly at random
from the set of all vertices. 
Key graphs have been studied in \cite{PMMPR08,BG09,YM13}. 

The theory of random graphs is a widely studied topic and many results are known. These results can be used to increase fault tolerance and connectivity, and decrease storage costs and communication costs of a random predistribution scheme.

It can be seen that the degree of the node signifies the number of nodes which share keys with this node, and which are within its communication range. 
To ensure that the time taken to transmit a message between any two nodes is low, 
the diameter and average path length of the graph should be low. 
Another important aspect is the connectivity of nodes during node compromise. 
In a network, there can be either node failure or communication link failure. 
How does the rest of the network behave? There are many interesting questions to answer while studying the fault tolerance or resilience of networks. 
One commonly used metric in key predistribution is the fraction of edges that are disconnected or the number of nodes 
that are isolated due to the compromise of a certain number of nodes. 
However, it does not entirely capture the notion of resilience. 
This is because, there might be a few components which are fully connected, thereby decreasing the fraction of edges disconnected. 
However, such a network is not fault-tolerant. 
A \emph{giant component} in a random graph contains a constant fraction of the total number of vertices of the graph. 
The size of the giant component is often a measure of resilience of a random graph. 
To ensure that messages are transmitted efficiently, it is important to study the average path length and diameter of a graph.

Keeping in mind the requirements of sensor networks, we propose a new model of key predistribution, which is based on the preferential attachment model. 
Preferential attachment (PA) model has its roots in Herbert Simon's 1955 paper \cite{S55}. 
The model was analyzed by Albert Barabasi and Reka Albert in \cite{BA02}. 
The basic idea is to grow the network in such a way that the overall global effect is the same. 
In the basic model, at each time step, a node is added and an edge is added to another node with a probability proportional to its degree. 
The degree distribution follows a power law, meaning that the probability that a node has degree $k$ is proportional to $k^{-\alpha}$, where $\alpha \ge 2$ is a constant.
The plot of the degree distribution has a long tail. 
There are a few nodes of very high degree and many nodes of low degree. 
In this way, a rich (high degree) node gets richer (more links). This phenomenon is often called the \emph{rich get richer} effect. 
Preferential attachment is widely observed in many real networks like social networks, citation networks, power grids, biological networks, and many more. 
The diameter of such graphs is $O(\log \log n)$, where $n$ is the number of nodes. 
This is in contrast to ER random graphs whose diameter is $O(\log n)$. 

Consider the following generative model for key predistribution. 
For each node $v$, we choose $k$ nodes following the preferential attachment model. 
A unique key is chosen for each of these $k$ nodes. 
Node $v$ shares a unique key with each of these $k$ nodes. 
In this way, nodes are assigned keys in such a manner that if there is a link between two nodes in the graph, then
there is a unique common key between the nodes. 
We assume full visibility, meaning that every node is within communication range of every other node. 

On the positive side, the networks are connected and have small average path length and small diameter, meaning that
any two nodes can communicate using only a small number of hops. 
In \cite{EG02}, average path length and diameter can be very long. 
In \cite{EG02}, the graph might not even be connected. 
However, in our scheme a few nodes will have very high degree, which means that the storage cost is substantially high for these nodes. 
In order to address this problem, we propose the \emph{preferential attachment model with degree bound (PA-DB)}, in which
edges are joined to a node with probability proportional to the degree of vertices, with an added restriction. 
A bound $d_{\max}$ is fixed on the maximum degree that the graph can have.
For every new node, a node is selected with probability proportional to the degree of those nodes, which have degree at most $d_{\max}-1$. 
This ensures that a node can have degree at most $d_{\max}$. 
This overcomes the problem of high storage cost for some nodes. 

Restricting the maximum degree of a graph is an important requirement in many applications. In peer-to-peer (P2P) networks, each node is a computer with limited system resources. Each new connection requires resources like CPU cycle, memory, disk space and network bandwidth. Hence, it is necessary to limit the number of connections that a node has to maintain, which effectively bounds the node's degree. An application in computing the streaming capacity in P2P networks is given in \cite{Liu2010}.

We study the following properties of this PA-DB graph: degree distribution, diameter and average path length. 
We also study fault-tolerance aspects such as the size of the giant component, when a few nodes fail. 
The construction when applied to predistribution results in storage and communication efficient schemes. 
Added to this, there are no shared key discovery phase, because each node knows its neighbor and the common key. 

This problem opens up a new direction of research. At a theoretical level, estimating different parameters of the PA-DB network
is a challenging task. There are many models for growing graphs under preferential attachment. 
So, it is interesting to see how this PA-DB model behaves. 
On a practical side, since networks in real life like social networks, sensor networks and power grids  follow power law distribution, this type of key predistribution can be used in such networks. 
In most cases like power grids, the capacity of a hub is limited, so preferential attachment with degree bound makes more sense than normal preferential attachment.

\subsection{Our Contributions}
\begin{enumerate}
\item We initiate the study of preferential attachment with degree bound, in which nodes can have degree at most $d_{\max}$, 
where $d_{\max}$ is chosen depending upon the application. 
\item We empirically study different parameters of the network formed by this model. 
In particular, we study the degree distribution, average path length, number of isolated nodes, number of edges removed, and size of the giant component.
\item A new key predistribution scheme is proposed using the above model. 
\item The predistribution scheme results in low storage and communication costs, low average path length, and high resilience compared to existing predistribution schemes. 
\item We open a new direction of research and pose many open problems. 
\end{enumerate}

\subsection{Organization}
The paper is organized as follows. In Section \ref{sec:related work}, we present related works for  
key predistribution. Mathematical background on preferential attachment is presented in Section \ref{sec:background}. 
Our model and its analysis are presented in Section \ref{sec:model}. In Section \ref{sec:KPD}, we 
study the predistribution scheme and compare it with existing schemes. 
We conclude in Section \ref{sec:conclusion} with future direction of research and some open problems.  

\section{Related Work}
\label{sec:related work}

Key predistribution (KPD) in WSN was first proposed by Eschenaur and Gligor (EG) \cite{EG02} as a practical lightweight alternative 
of key management in resource constrained sensors. 
As discussed in the introduction, their scheme was probabilistic in nature, and did not guarantee connectivity of the network. 
Finding common key is also difficult and required communication overhead of $O(k\log n)$, as shown in \cite{RR08b}. 
Chan, Perrig and Song (CPS) \cite{CPS03} proposed a pairwise scheme, in which each node shares a unique common key with 
at least $k$ nodes selected at random. 
This ensures full resilience, meaning that even if some nodes are compromised, the rest of the key sharing links remain unaffected. 
On the flip side, some nodes can have arbitrarily large number of keys and some nodes will have few keys. 
There is no guarantee that the network will be connected. 

Thereafter, many key predistribution schemes have been proposed, which trade-off storage with resilience and connectivity. 
Combinatorial schemes like \cite{CY04,LS04,RR07} are a class of deterministic schemes which were very popular because of their simplicity of construction and
simple algorithms to find key sharing neighbors. However, such schemes do not scale well and have poor resilience. 
Camtepe and Yener \cite{CY04} were the first to propose such schemes using combinatorial designs called \emph{projective planes}. 
Lee and Stinson \cite{LS04} proposed KPD schemes using transversal designs, while Ruj and Roy \cite{RR07,RR08c} used partially
balanced incomplete block designs and Reed-Solomon codes.
In Section \ref{sec:KPD}, we will show that the Lee-Stinson (LS) scheme \cite{LS04} has very poor resilience compared to random key predistribution schemes. 
A survey of such schemes appear in \cite{XRSDHG07,RNS11}. 

Most of these schemes studied the fraction of links compromised and number of nodes isolated under node compromise. 
However, these are not the only measures of resilience. 
Pietro \emph{et al.} \cite{PMMPR08} showed that connectivity via secure links and resilience against malicious attacks can be achieved
simultaneously in random EG scheme. 
They showed that an adversary cannot partition the network into two components of linear size, unless they compromise linearly many nodes. 
These networks are \emph{unsplittable} with high probability, and \emph{redoubtable}, meaning that even by compromising a large number of nodes, 
the confidentiality is not lost. 

Blackburn and Gerke \cite{BG09} studied \emph{uniform random intersection graphs}. The problem is to attach a list of $k$ colors to 
each of $n$ nodes, the colors being chosen uniformly at random from a set of $m$ colors, such that
two nodes are connected if they have a common color in their list. It can be seen that EG network is an example of these types of graphs. 
They analyzed the connectivity of such graphs. 

Yagan \emph{et al} \cite{YM08,YM13,YM12} studied the key graphs formed by \cite{EG02} and \cite{CPS03} under full and partial visibility. 
They showed that graphs defined in \cite{CPS03} are connected with high probability. 
They also studied the security under the ON-OFF secure channel, which means that the links between nodes  might or might not be active at a given instant. 

\section{Background}
\label{sec:background}

Preferential attachment (PA) is a well studied model for generating random graphs. 
The main idea behind preferential attachment is that, at each step a new node is added with an edge with the other 
end point chosen from the set of existing nodes with a probability proportional to the degree of the node. 
This implies that a high degree node has a higher chance of being chosen. 
There are a large number of nodes with low degree and a few nodes (also called \emph{hubs}) with very high degree. 
The degree distribution of the nodes follow a power law, meaning that the probability that a node has degree $k$ is given by $P(k) \propto k^{-\alpha}$, 
where $\alpha \geq 2$ is a constant. 
The plot of $P(k)$ against $k$ gives a curve with a long tail. 

These graphs are known to possess a \emph{scale free property}, meaning that 
the degree distribution of nodes still follow a power law with the same coefficient $\alpha$, even when sampling at different time intervals. 
In other words, if instead of one edge $k$ edges are selected with probability proportional to the degree, the same degree distribution is obtained. 

Scale-free networks are popular because they occur in many real life situations. For example, 
in a social network, a person who is already popular has a higher probability of acquiring new friends. 
Scale-free networks also occur in citation networks, power grids, biological networks and many others. 

These are the parameters for the PA model \cite{CL10}: 
\begin{enumerate}
\item Number of nodes $n$ in the final graph $G$,
\item A probability $p$, $0\leq p\leq1$,
\item An initial graph $G_0$ at time $t=0$. 
\end{enumerate}
Generally, we take $G_0$ to be a single vertex with a self loop. 

The random graph $G_t(p,G_0)$ is constructed as defined below:

\begin{enumerate}
\item Begin with the initial graph $G_0$.
\item At time $t>0$, the graph $G_t$ is formed from $G_{t-1}$ as follows:
\begin{enumerate}
\item With probability $p$, choose a vertex $v \in V(G_{t-1})$ uniformly at random.
\item With probability $1-p$, choose a vertex $v \in V(G_{t-1})$ randomly with probability proportional to its degree $deg(v)$.
\end{enumerate}
\end{enumerate}

The degree distribution for the random graph $G_t$ follows a power law. The probability that a node has degree $k$ is given by $P(k) \propto k^{-\alpha}$, where,  
$\alpha = 1+\frac{1}{1-p}$. 
Since $\sum_{k=1}^{\infty}P(k) =1$, $P(k)= \frac{1}{\zeta(\alpha)}k^{-\alpha}$, where $\zeta(t)$ is the \emph{Riemann zeta function}, and is given by 
$\zeta(t) = \sum_{k=1}^{\infty} k^{-t}$.
For detailed study of the model, one can refer to \cite{EK10,CL10}. 

\begin{algorithm}[ht]
\caption{$PA-DB(G_0,n,p,d_{\max},k)$} \label{PADB}
\begin{algorithmic}[1]
\STATE $V(G_0) \leftarrow \{v_0,\ldots,v_{k-1}\}$;
\STATE $E(G_0) \leftarrow \{(v_0,v_0),\ldots,(v_{k-1},v_{k-1})\}$;
\FOR{$t = k,\ldots,n-1$}
\STATE Create a new vertex $v_t$;
\STATE Choose $r$ uniformly at random from the interval $[0,1]$;
\IF{$r \le p$}
	\STATE Choose $k$ distinct nodes (without replacement) $\{u_1,\ldots,u_k\}$ uniformly at random from those vertices in $V(G_{t-1})$ whose degree is at most $d_{\max}-1$;
\ELSE
	\STATE Choose $k$ distinct nodes (without replacement) $\{u_1,\ldots,u_k\}$ with probability proportional to $deg(u_i), 1 \le i \le k$ from those vertices in $V(G_{t-1})$ whose degree is at most $d_{\max}-1$;
\ENDIF
\STATE $V(G_{t}) = V(G_{t-1}) \cup \{v_t\}$;
\STATE $E(G_{t}) = E(G_{t-1}) \cup \{(v_t,u_1), \ldots, (v_t,u_k)\}$;
\ENDFOR
\RETURN $G_{n-1}$;
\end{algorithmic}
\end{algorithm}

\begin{table*}[ht]
\label{table:path-dia}
\caption{Average path length and diameter of graphs generated using PA and PA-DB models}
\begin{center}
\begin{tabular}{|c|c|c|c|c|c|}
\hline
\textbf{Metric} & \textbf{PA} & \multicolumn{4}{|c|}{\textbf{PA-DB}} \\
\cline{3-6}
& & $d_{\max} = 15$ & $d_{\max} = 20$& $d_{\max} = 25$ & $d_{\max} = 30$\\
\hline
\textbf{Average path length} & 3.36 & 3.70 & 3.61 & 3.55 & 3.46 \\
\hline
\textbf{Diameter} & 5 & 7 & 6 & 6 & 6 \\
\hline
\end{tabular}
\end{center}
\end{table*} 

\section{Proposed model of preferential attachment with degree bound}
\label{sec:model}
We first present Algorithm \ref{PADB} for preferential attachment with degree bound, and then show how it compares with the traditional preferential attachment model. The initial graph $G_0$ consists of $k$ isolated nodes $v_0,\ldots,v_{k-1}$ and each node $v_i$ has a self-loop $(v_i,v_i)$.

The parameters to be considered are:
\begin{enumerate}
\item An initial graph $G_0$ at time $t=0$,
\item Number of nodes $n$ in the final graph $G$,
\item A probability $p$, $0\leq p \leq 1$, 
\item A degree bound $d_{\max}$.
\item Number of neighbors (edges) $k$ added to a new node.
\end{enumerate}

\begin{figure}[ht]
\begin{centering}
\includegraphics[width=3.0in]{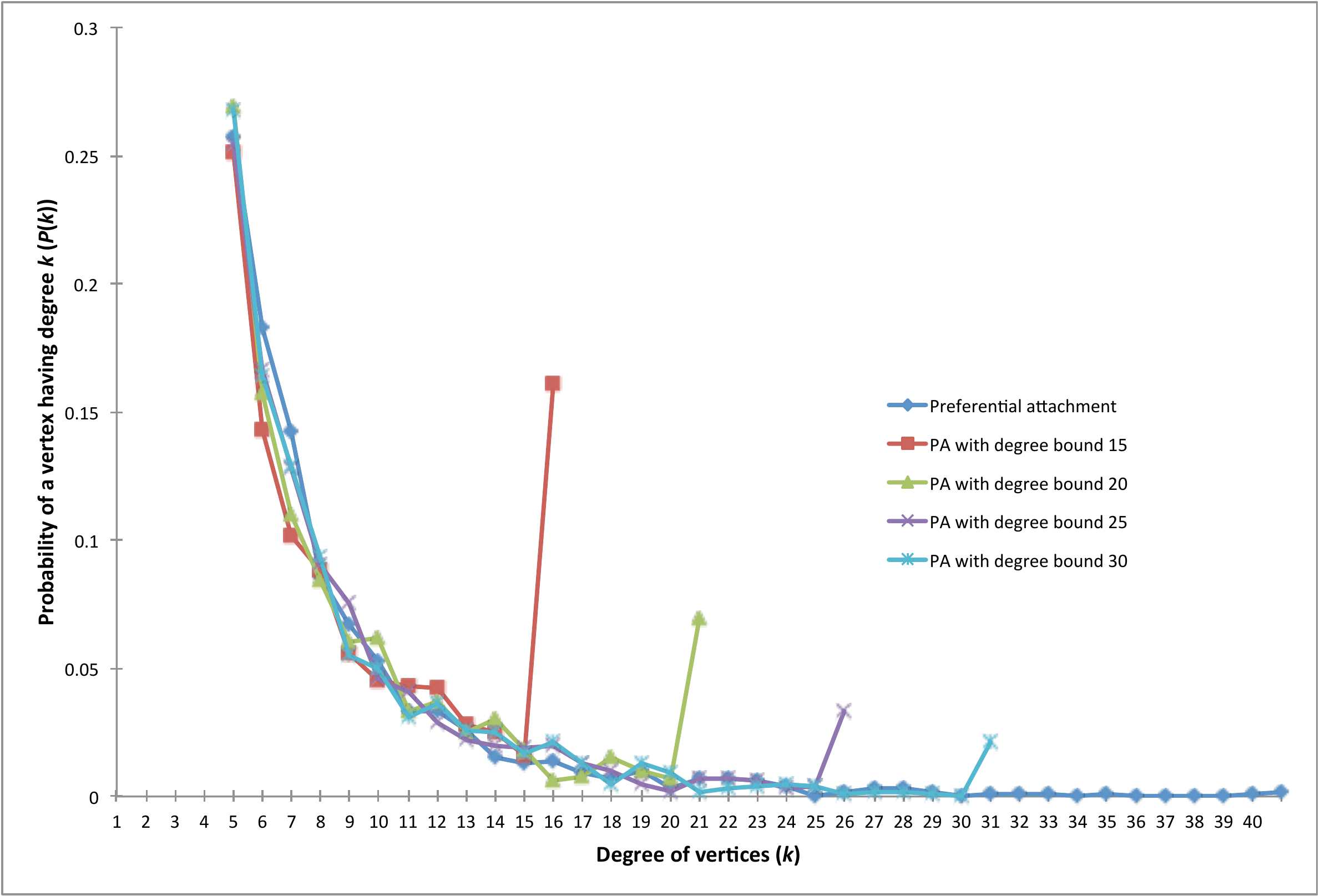}
\caption{Degree Distribution for PA with and without degree bound}
\label{fig:PA-comp}
\end{centering}
\end{figure}

In Algorithm \ref{PADB} for preferential attachment with degree bound, instead of connecting every new vertex to only one existing vertex, it is being connected to $k$ other vertices. 
This results in a graph with degree at least $k$. Such graphs will be used in our predistribution scheme in Section \ref{sec:KPD}.  



We consider graphs on $n = 1000$ vertices with $p=0.4$ and $k=4$ edges attached to every new vertex at each step. 
Therefore $\alpha = 1 + \frac{1}{1-p} = 2.67$ for the PA model.
The following graphs are compared in Fig.  \ref{fig:PA-comp}:
\begin{enumerate}
\item Normal PA model. 
\item PA with degree bound $d_{\max} = 15, 20, 25, 30$.
\end{enumerate}

For the PA-DB model, the number of nodes with degree equal to $d_{\max}$ is significantly higher than nodes of degrees $d_{\max}-1, d_{\max}-2, \ldots$. 
One reason is that the nodes which could have degree greater than $d_{\max}$ now have a degree of $d_{\max}$.

We compare the average path length and the diameter of the graphs in Table I.
The average path length is slightly longer for PA-DB than PA. 
For a network with 1000 nodes, $k=4$, degree bound 15, and $p=0.4$, the average path length is 3.70 and the diameter is 7. 
Without degree bound, the average path length is 3.36 and the diameter is 5. 

\section{New Key predistribution scheme using PA-DB}
\label{sec:KPD}

We show how to use preferential attachment with degree bound for key predistribution. 
Let $G(n,p)$ be a graph obtained by executing Algorithm \ref{PADB} given in Section \ref{sec:model}. Observe that at each step, a new vertex is connected to $k$ other vertices. 
Thus, each node has at least $k$ edges incident to it.  
 For simplicity,  we refer to the graph as $G$. 
Each vertex in $G$ corresponds to a node in the network.
The set of neighbors of a vertex $v \in V(G)$ is denoted by $nbd(v)$. 
For each sensor node $v$, we assign $|nbd(v)|$ unique pairwise keys, one for each $u \in nbd(v)$. 
The closest model is \cite{CPS03}, where each node selects $k$ nodes at random and share a unique pairwise key with each of the $k$ nodes. 

A node $v$ thus maintains the identity of its neighbors and the unique common key shared with each of the neighbors. 
The degree bound ensures that each node stores at most $d_{\max}$ keys. 
The high degree nodes can be considered as cluster heads in a hierarchical network. 

We now check the various properties of this key predistribution scheme. 

\begin{enumerate}
\item There is a vertex for each sensor node in the network. Thus there are $n$ nodes in the network.
\item The degree bound ensures that each node stores at most $d_{\max}$ keys.
\item The network is fully connected when no nodes are compromised (by the construction of the graph). 
\item The diameter of the graph is $O(\log\log n)$. This means that the number of message transmission is small, compared to the size of the network.
\item Since a node shares pairwise keys with other nodes to which it is connected, it is fully resilient to node compromise. 
This means, that apart from the nodes that are compromised, no other links between the existing nodes are broken. 
\end{enumerate}

We now consider the effect of node compromise on a wireless sensor network (WSN). 

\subsection{Adversarial model}
The adversarial models generally considered in WSN are random node compromise and targeted node compromise. 
When a node is compromised, an adversary can read all the information stored in the node, including the data and the keys. 
The main problem is how to prevent the adversary to receive any information about the data or keys stored in other nodes. 
In most key predistribution schemes, the set of keys belonging to the non compromised nodes can be calculated. 
In the scheme given in \cite{EG02}, the nodes exchange the identifiers of the keys, so a passive adversary knows which key identifiers are present in which node. 
Thus, an adversary can selectively compromise nodes, such that the size of the set of keys compromised is maximized. One way of doing this, 
is to select nodes, which have disjoint set of key identifiers or nodes where the set of common key identifiers is few. 
In most of the combinatorial schemes, the nodes exchange the identities of the nodes, and calculate the common key identifiers.
Since the key identifier algorithm is known publicly, an adversary knows which node has what set of keys, though nodes do not exchange them 
explicitly. 
Thus, selective node compromise is possible in this case as well. 

In our proposed model, a node only shares unique common keys with a set of nodes. Only if an attacker compromises a node, does it know
which node contains what common keys with which other nodes. 
The key distribution algorithm speaks nothing about the assignment of keys to nodes. 
This is because, a node only know its neighbors and the common keys. Even if the key generation algorithm above is made public, 
the node identifiers might not be sequentially assigned. 

When nodes are deployed in the field, the structure of all the nodes are similar. 
Messages are sent intermittently, so it is not easy to do a traffic analysis and find out a high degree or high betweenness node (\emph{betweenness} of a node is the number of shortest paths through that node). 
Thus, selective node compromise by capturing high degree nodes or high betweenness nodes do not arise for this application. 

\subsection{Resilience under node compromise}
We consider only random node compromise. 
An adversary can select and  compromise $s$ nodes at random. 
We denote the fraction of nodes compromised by $fs$. Thus $fs = s/n$. 
We study the resilience of the network by comparing the following parameters:

\begin{enumerate}
\item Fraction of nodes isolated when $s$ nodes are compromised. 
We denote this metric by $V(s)$. 
\item Fraction of edges removed when $s$ nodes are compromised.  
We denote this metric by $E(s)$. 
\item The size of the largest component. We denote the fraction of nodes in the largest component by $C(s)$. 
\item The average path length $P(s)$ in the resulting graph after $s$ nodes are compromised. 
\end{enumerate}

The first two measures have been widely used in the literature \cite{RR07,RNS11,RNS13}. 
We propose the third and the fourth measures, because they capture fault tolerance quite well. 
For example, there might be a few components each forming a clique, thereby increasing the number of edges in the resulting graph. 
According to the previous measures, although the graph might look to be highly resilient, in reality it is not. 

\subsection{Experimental setup}
We now discuss each of the metrics in details and show how our results compare with that of existing schemes. 
There are 10000 nodes in each network. The nodes are deployed randomly. 

We compare our scheme with that of: 
\begin{enumerate}
\item Eschenaur and Gligor (EG) scheme \cite{EG02}.
\item CPS scheme \cite{CPS03}, which gives a $k$-out graph.
\item Combinatorial scheme of Lee and Stinson (LS) \cite{LS04}.
\item Combinatorial scheme of Ruj \emph{et al.} (RNS) \cite{RNS13}. 
\end{enumerate}

For CPS scheme, we choose $k = 7$. We observe that maximum degree of a node is 25. 
For EG scheme, we choose the size of key chain as $25$ and a key pool of 500000 keys. 
We choose LS and RNS scheme with $k=25$. For our scheme, we choose $d_{max}=25$.
There are many other schemes present in the literature. For example, there are many deterministic schemes. However, we
do not consider all of them, because deterministic schemes perform poorly as the number of compromised nodes increases. 
The specific pattern also limits the scalability of the networks. 

\begin{figure}[ht]
\begin{centering}
\includegraphics[width=3.0in]{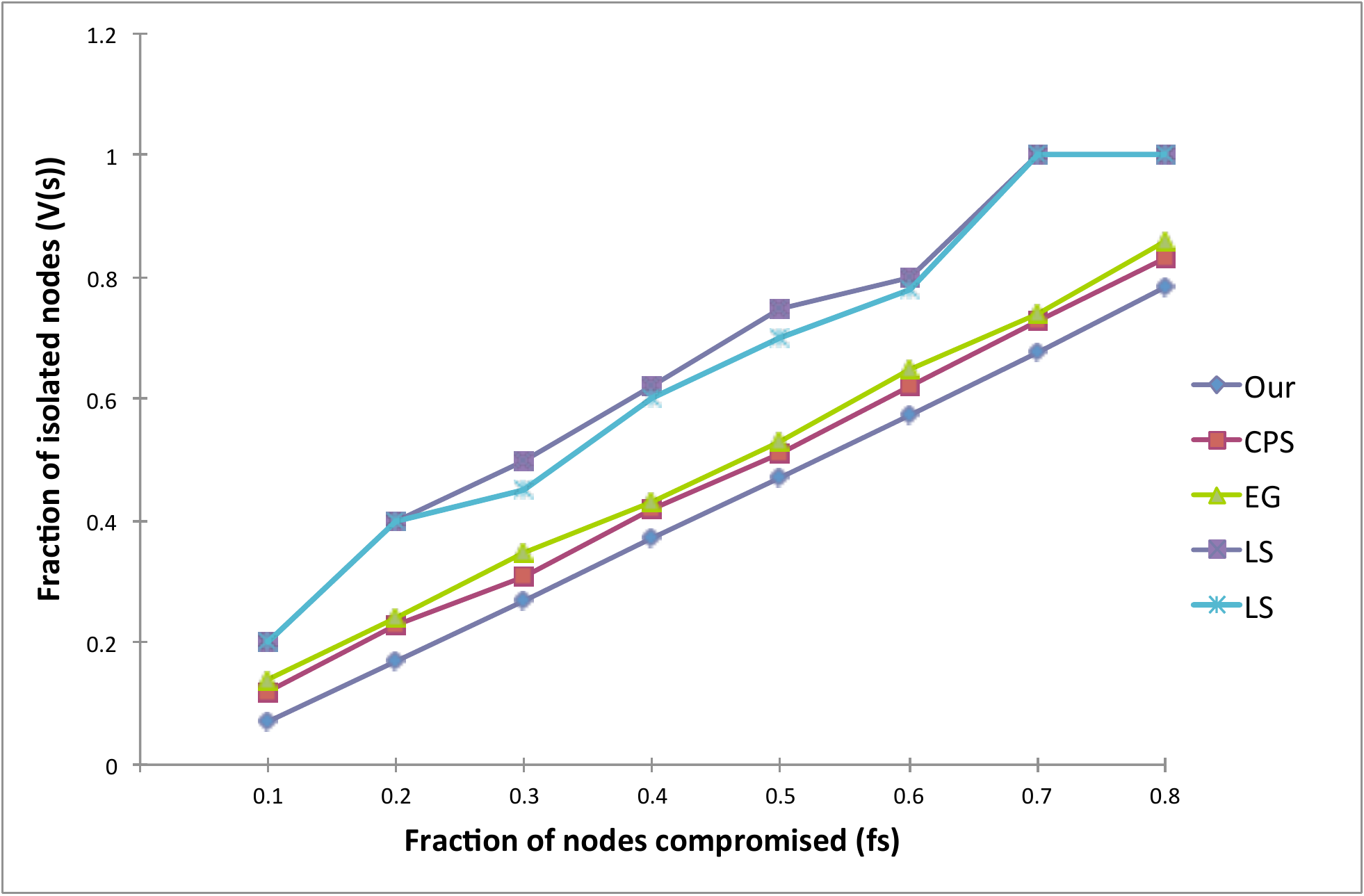}
\caption{Comparison of fraction of vertices isolated  when fraction of $fs$ nodes are compromised}
\label{fig:V_s}
\end{centering}
\end{figure}

\begin{figure}[ht]
\begin{centering}
\includegraphics[width=3.0in]{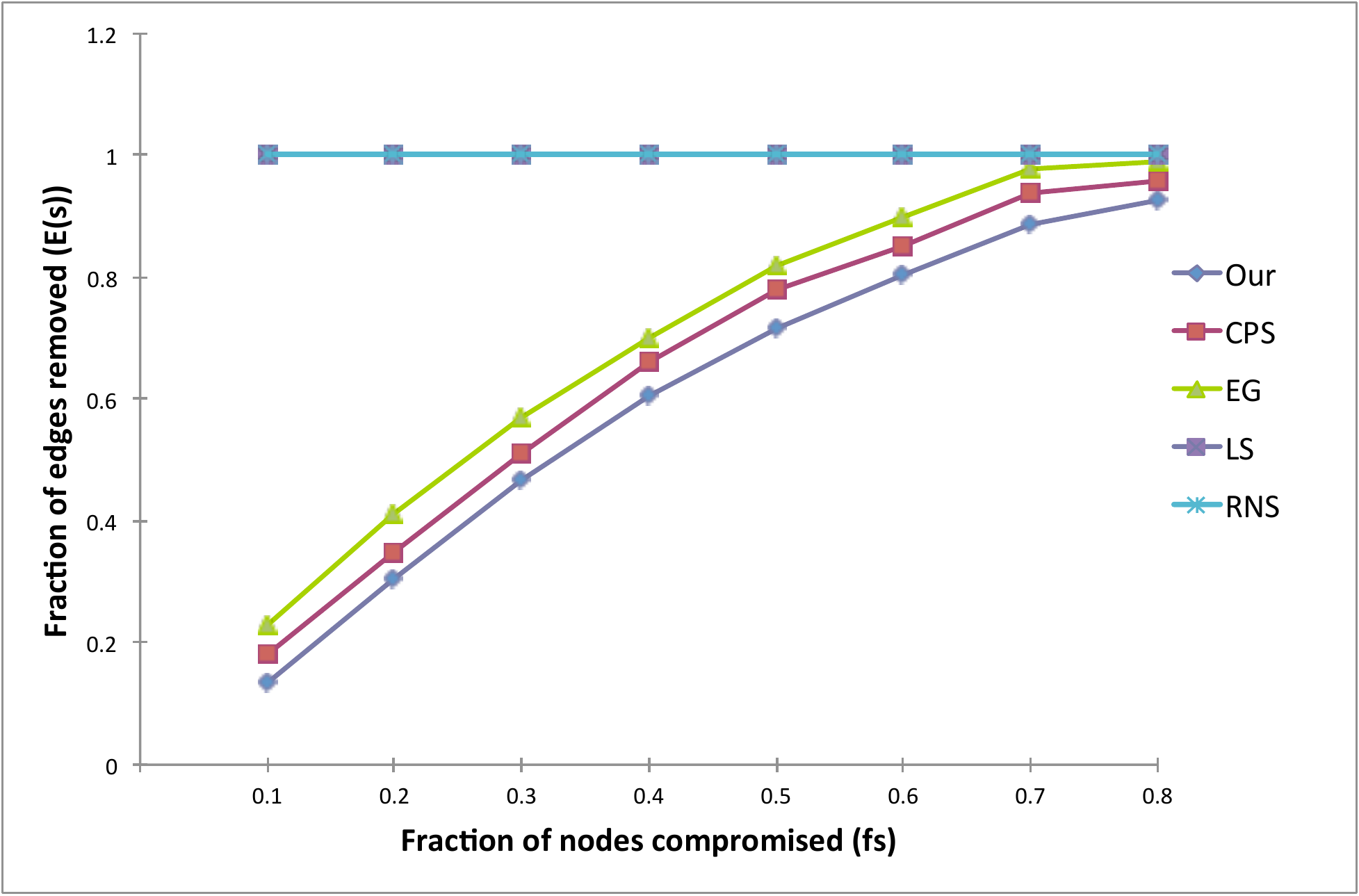}
\caption{Comparison of fraction of edges removed  when fraction of $fs$ nodes are compromised}
\label{fig:E_s}
\end{centering}
\end{figure}

\begin{figure}[ht]
\begin{centering}
\includegraphics[width=3.0in]{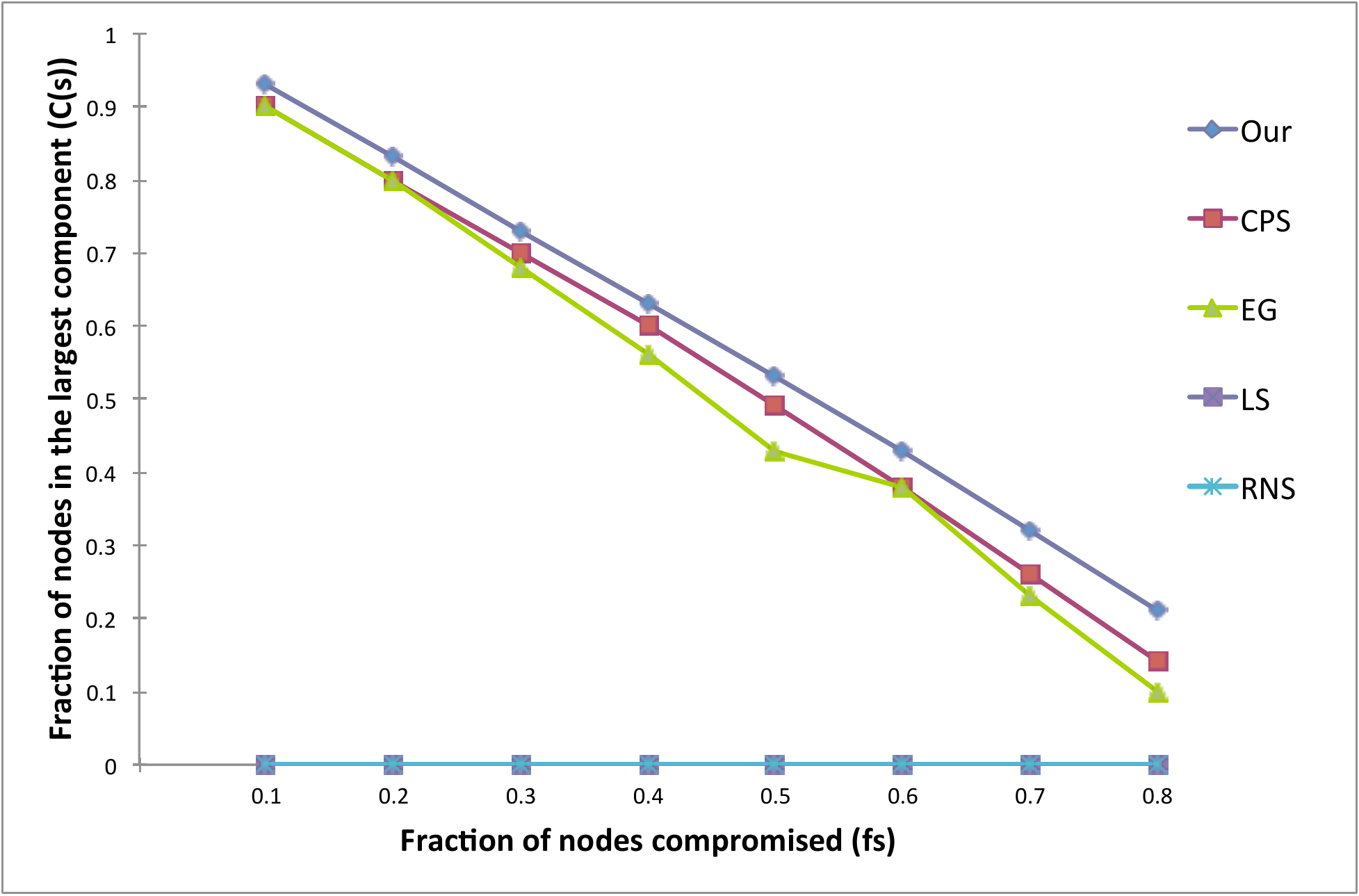}
\caption{Comparison of fraction of nodes in the largest component when fraction of $fs$ nodes are compromised}
\label{fig:C_s}
\end{centering}
\end{figure}

\begin{figure}[ht]
\begin{centering}
\includegraphics[width=3.0in]{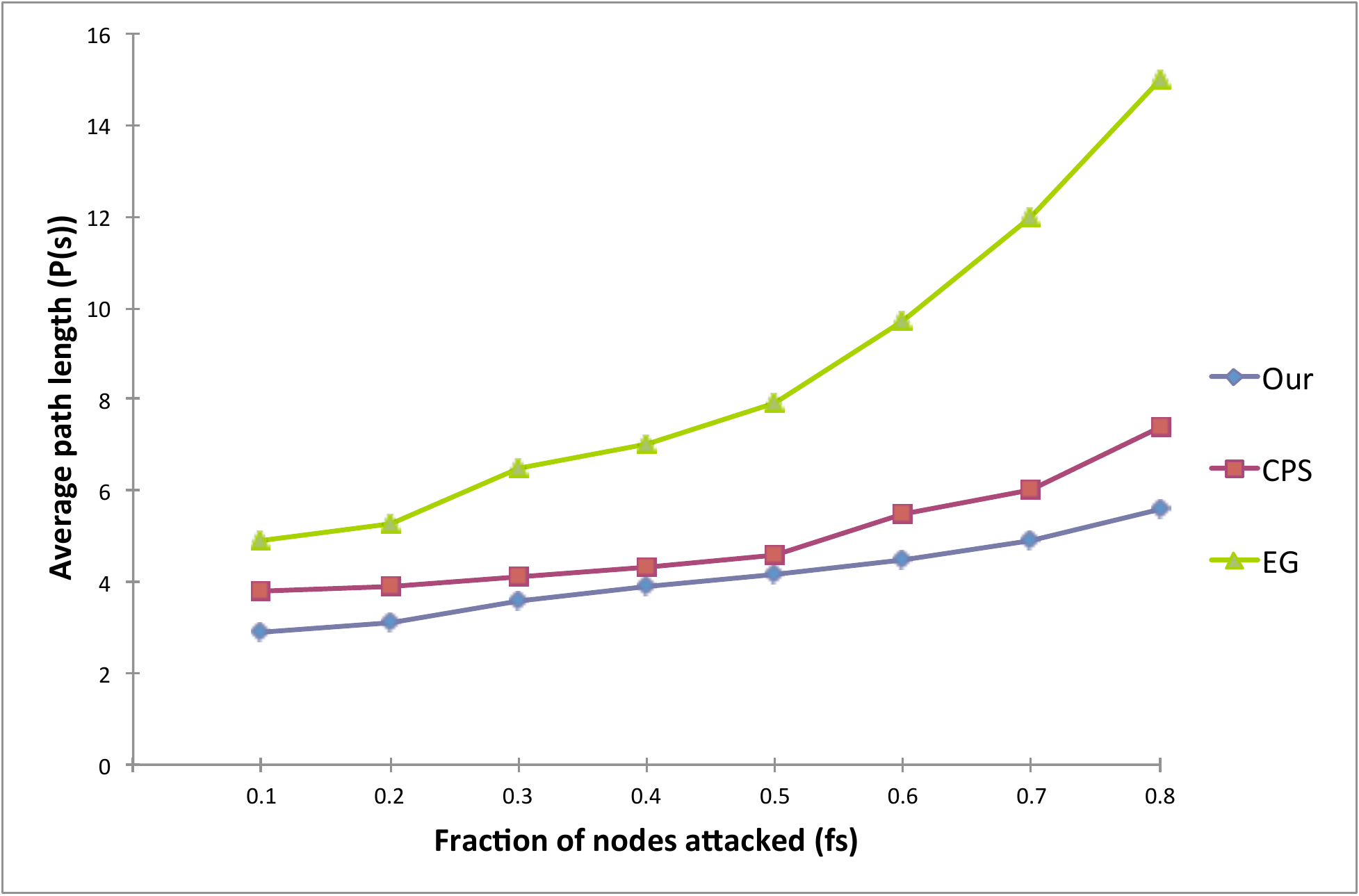}
\caption{Comparison of Average path length when fraction of $fs$ nodes are compromised}
\label{fig:P_s}
\end{centering}
\end{figure}

\subsection{Fraction of nodes isolated}

We define $V(s)$ to be the ratio of the number of nodes isolated and the total number of nodes in the network. 
We compare the $V(s)$ values in Fig \ref{fig:V_s}, when a constant fraction of nodes are compromised. 
Our scheme performs best among the four schemes. The LS scheme performs worst, followed by the RNS scheme, EG scheme and the CPS scheme. 

\subsection{Fraction of edges removed}

The total number of edges removed is bounded by the number of  edges that are incident on the compromised nodes. 
If we consider the sum of the degrees of all the compromised nodes, each removed edge is counted either once or twice. 
It is counted twice, if both the end points are compromised and once if only one end point is compromised. 
Our scheme is fully secure, meaning that when $s$ nodes are compromised, 
none of the edges between any two uncompromised nodes are broken. 
This feature is also available in CPS scheme.

In Fig \ref{fig:E_s}, we compare our scheme with that of LS, EG and CPS. We note that our scheme has the lowest value of $E(s)$. 
CPS and EG has comparable but higher $E(s)$, whereas LS and RNS have the worst $E(s) \approx 1$. 

\subsection{Fraction of nodes in the largest component}
We look at the graph formed by the key predistribution scheme. When nodes are compromised, 
nodes and edges will be broken. Thus, the graph can disintegrate into many components. 
From our experiments, we observe that the network of LS scheme disintegrates into many components, followed by EG and CPS. 
In our scheme there are very few components, and it has been seen that most of these components consist of isolated vertices. 
From Fig. \ref{fig:C_s}, the size of the largest component is highest in our scheme and almost 0 in case of LS and RNS. This means that the graph disintegrates
into small components and there are no giant components in LS. The values of $C(s)$ for the CPS and the EG schemes lie in between.

\subsection{Average path length}
We have compared the path lengths of various schemes in Fig. \ref{fig:P_s}. The average path length is smallest in our scheme. The corresponding value for CPS is larger and that in EG the largest. Thus, in our scheme, messages can be transmitted quickly from one node to other nodes. 
We have not considered the LS and RNS schemes, because the graph disintegrates when nodes are compromised. 

\section{Conclusion and open problems}
\label{sec:conclusion}
We have proposed a new model of preferential attachment. Here the maximum degree is upper bounded by $d_{\max}$. 
The new model is applied to key predistribution in wireless sensor networks and it is found to be better than existing key predistribution schemes. 
We have seen that PA with degree bound has applications in many real life scenarios like power grids, where nodes have limited capacity. We can also think of the highest degree nodes as cluster heads or hubs. 
A hub cannot have a very large degree. 
We believe that this model is more suitable for IoT and cyber-physical systems than the conventional PA model. 

There are many interesting open problems. We list a few of them below.
\begin{itemize}
	\item Mathematical analysis of this model to find the average path length, diameter, betweenness, degree distribution, size of the giant component, clustering coefficient and other graph parameters.
	\item What is the effect of the maximum degree bound $d_{\max}$ on the degree distribution of the graph?
	\item There are many PA models. It is interesting to see the effect of degree distribution with respect to these models.
	\item Can we vary $d_{\max}$ to make the graph more fault-tolerant under random attacks?
	\item In a more realistic scenario, instead of a single degree bound $d_{\max}$ for all vertices, we have a degree bound $d_v$ for every vertex $v$. Our model is a special case of this model. Study of such models is interesting in its own right.
\end{itemize}


\bibliographystyle{plain}
\bibliography{sensor,smartgrid}
\end{document}